# Remanufacturing cost analysis under uncertain core quality and return conditions: extreme and non-extreme scenarios

Saeed Z.Gavidel[1], Jeremy L.Rickli[2]

[1,2]*Industrial and Systems Engineering, Wayne State University, Detroit, Michigan, USA*

Saeed Z.Gavidel is the corresponding author for this submission. Mailing and email addresses are included for each of the authors.

[1]Department of Industrial and Systems Engineering
Wayne State University
4815 Fourth Street, Rm 1062
Detroit, MI 48202
Phone: (586) 945-0021
szgavidel@wayne.edu

[2]Department of Industrial and Systems Engineering
Wayne State University
4815 Fourth Street, Rm 2173
Detroit, MI 48202
Phone: (313) 577-1752
jlrickli@wayne.edu

# Remanufacturing cost analysis under uncertain core quality and return conditions: extreme and non-extreme scenarios

Uncertainties in core quality condition, return quantity and timing can propagate and accumulate in process cost and complicate cost assessments. However, regardless of cost assessment complexities, accurate cost models are required for successful remanufacturing operation management. In this paper, joint effects of core quality condition, return quantity, and timing on remanufacturing cost under normal and extreme return conditions is analyzed. To conduct this analysis, a novel multivariate stochastic model called Stochastic Cost of Remanufacturing Model (SCoRM) is developed. In building SCoRM, a Hybrid Pareto Distribution (HPD), Bernoulli process, and a polynomial cost function are employed. It is discussed that core return process can be characterized as a Discrete Time Markov Chain (DTMC). In a case study, SCoRM is applied to assess remanufacturing costs of steam traps of a chemical complex. Its accuracy analyzed and variations of SCoRM in predictive tasks assessed by bootstrapping technique. Through this variation analysis the best and worst cost scenarios determined. Finally, to generate comparative insights regarding predictive performance of SCoRM, the model is compared to artificial neural network, support vector machine, generalized additive model, and random forest algorithms. Results indicate that SCoRM can be efficiently utilized to analyze remanufacturing cost.

# 1. Introduction

Remanufacturing is the ultimate form of recycling and considered by some researchers to be the best practice of circular economy (Steinhilper, 1998; Fadeyi et al., 2017). Remanufacturing is defined as: "The process of returning a used product to at least OEM original performance specification from the customers' perspective and giving the resultant product a warranty that is at least equal to that of a newly manufactured equivalent." (Bhamra and Hon, 2004). Used-products, also known as End-of-Use (EoU) cores, are material source of remanufacturing processes. However, in general, core quality condition, return quantity, and timing can be highly uncertain and this uncertainty complicates remanufacturing operations like Product Acquisition Management (PrAM) and production planning (Ilgin and Gupta, 2012). Guide (2000) categorized these uncertainties as top challenges in remanufacturing. In a Strengths, Weaknesses, Opportunities, Threats (SWOT) analysis, the uncertain nature of core quality, return quantity, and timing are reported as weakness for remanufacturing processes (D'Adamo and Rosa, 2016).

In passive acquisition programs, remanufacturers must admit all returning cores regardless of quality condition, return quantity and timing (Guide, 2000). Obligations like Extended Producer Responsibility (EPR) mandate manufacturers to passively acquire EoU cores (Organisation for Economic Co-operation and Development, 2001). Passive core acquisition can also be an internal policy of a company. For example, consider returns of industrial valves to a valve repair shop of a chemical complex where, regardless of quality, quantity, and timing of returns, the repair shop must admit all returns. Repair is an equivalent term for remanufacture in valve industry (Hauser and Lund, 2008). Systems with passive core acquisition programs (Fig. 1), are under the risk of extreme returns (Guide, Van Wassenhove, 2001).

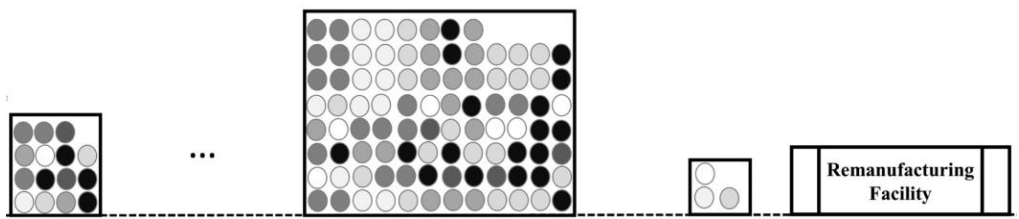

Figure 1. Passive core return process

In passive returns, EoU cores return at random time periods in batches of random size (number of cores in batch) that can be extreme, where each batch contains cores with random quality condition, illustrated by different shades in Fig.1. Core quality, return quantity, and timing are uncertainty sources transmitted to remanufacturing systems jointly, hence; they should be analyzed jointly. For the case of cost models, uncertainties in core quality condition, return quantity, and timing can collectively propagate and accumulate in remanufacturing cost. Therefore, disjoint analysis of these three uncertain factors can yield inefficient cost models. This inefficiency can be even more pronounced if extreme returns are taken into account.

In remanufacturing, cost models are used extensively for different purposes at strategic and operational levels. For instance, Robotis et al. (2012) used a linear cost model to study the necessity of strategic investment in product reusability in a hybrid manufacturing/remanufacturing system. At the operational level, Galbreth and Blackburn (2006) used a cost model to sort cores before remanufacturing. Profitability of a remanufacturing process at Pitney-Bowes, a global technology company well-known for its mailing equipment and services, was investigated by Ferguson et al. (2009) by developing a polynomial cost model. In an investigation to optimize reliability and process cost in remanufacturing, Jiang et al. (2016) developed a linear cost model to assess cumulative cost of remanufacturing by evaluating operational costs like machining and labor costs.

Considering the critical role of cost models in remanufacturing operations, more accurate and realistic cost models are desirable. This work is dedicated to answering the following research question: How core quality condition, return quantity and timing, as three uncertain factors, jointly influence remanufacturing costs under normal and extreme core return scenarios? To answer this question, a novel multivariate stochastic cost model, called the Stochastic Cost of Remanufacturing Model (SCoRM), is developed and applied to assess joint impacts of quality, quantity, and timing on remanufacturing costs under normal and extreme return conditions.

In the proposed SCoRM, the quantity of returns is modeled by a mixture Hybrid Pareto Distribution (HPD). Then, using the threshold of HPD, returned batches are classified into normal and extreme classes. By this classification, the return process is modeled as a Bernoulli process where returns timing follow the Geometric distribution. Thus, the core return process can be characterized as a two-state Discrete Time Markov Chain (DTMC) system. Finally, three uncertain factors, *i.e.*, core quality condition, return quantity, and timing are coupled with a polynomial cost function adopted from Fergusson and Guide (2006) to build the SCoRM. Accumulated uncertainties present in remanufacturing cost and originating from uncertainties in core quality, return quantity and timing, are evaluated by bootstrapping technique.

With SCoRM, decision makers have assistance in evaluating remanufacturing operations like PrAM, inventory planning, and budget management using resulting cost paths. A process cost path is a cost profile of the remanufacturing system where the cumulative sum of remanufacturing costs is mapped over the timespan of the process. For example, process cost paths can be used to evaluate the best and worst case cost scenarios. This investigation can also be used in PrAM operations where accurate prediction of remanufacturing costs supports core pricing decisions. Table 1 shows some potential applications of SCoRM.

Table 1. Potential applications of SCoRM

| Scope | Application |
|---|---|
| Cost-benefit Analysis | More accurate and realistic cost analysis |
| Risk Management | Risks of extreme returns can be analyzed |
| Budget Management | Assessing best and worst scenarios for remanufacturing costs |
| PrAM | Core pricing by considering remanufacturing costs |
| Human Resources | Considering short-time staffing to mitigate risks of extreme returns |
| Inventory Management | Capacity planning for inventory system in extreme returns |
| Core Preprocessing | Core sorting, prioritization/triage based on remanufacturing costs |

For the case of remanufacturing systems with active core acquisition paradigm, where remanufacturers have partial control on variations of quality, quantity and timing (Choi and Cheng, 2011), joint analysis of returns quality, quantity and timing can also be critical. For instance, consider the case of obtaining EoU cores from third party brokers according to as-needed model. In this active core acquisition model; 1) Brokers share risks with remanufacturers by premium charging (Guide, 2000), 2) since brokers own the cores, they can control the market, and 3) from a systematic perspective, the original issue of uncertainty in returns is not resolved/mitigated but is transmitted to brokers as another member of circular economy.

The remainder of this paper is organized as follows; Section 2 reviews existing cost models in remanufacturing literature. In Section 3, SCoRM is developed by building quantity, timing, and quality models, and joining these models through a polynomial cost function. In Section 4, SCoRM is used to analyze remanufacturing costs of steam traps returning to a valve shop of a chemical complex. Process cost paths for remanufacturing are generated and associated variations are evaluated by bootstrapping. Finally, the predictive capability of SCoRM is compared to multiple other predictive algorithms. Section 5 concludes that SCoRM can be efficiently used to evaluate the remanufacturing costs under uncertain quality, quantity and timing conditions where return quantity can be extreme.

## 2. Review of literature

In this section, existing cost models in remanufacturing literature are reviewed and compared with SCoRM. The models have been compared using two criteria; 1) model completeness: are joint impacts of quality, quantity, and timing on remanufacturing considered in developing the model? and 2) extreme analysis: is the model capable of handling both normal and extreme return scenarios?

In remanufacturing, different cost models have been developed for a variety of purposes. For example, to confirm the need of core sorting in remanufacturing, Galbreth and Blackburn (2006) used an s-shaped cost function to sort cores. They used a cutoff cost (maximum cost acceptable to justify remanufacturing) to sort cores into two classes, remanufacture and scrap. By assuming acquisition on as-needed basis, uncertainty in quantity and timing is assumed to be reduced. However, severity and impacts of unremoved uncertainties is unknown and extreme scenarios are not integrated into cost analysis. Robotis et al. (2012) also used a cutoff-based cost function to sort EoU cores into remanufacturable versus non-remanufacturable classes to analyze impacts of quality uncertainty on investments in product reusability. A core is assumed to be remanufacturable if the associated remanufacturing cost is less than manufacturing a new product. By this assumption, core return quantity is characterized by the Binomial distribution. Also, they confirmed that remanufacturers can achieve economy of scale if core returns is extreme, however, like Galbreth and Blackburn (2006), analysis of extreme scenarios is not considered.

A polynomial cost function was developed by Ferguson et al. (2009) to study the problem of tactical production planning of remanufacturing operations in Pitney-Bowes. They used this model to study impacts of quality-based core sorting on remanufacturing profitability. They showed that sorting improves remanufacturing profitability by average 4% and that this figure increases if return quantity increases. However, it is unknown if there is a desirable limit for return quantity before risks of extreme returns diminish advantages of increased returns. For example, increased return quantity will increase need for inventory and inventory costs will increase. It is indicated that the Pitney-Bowes leasing system reduced uncertainties in returns quantity and timing. However, it is not known if unremoved uncertainties are negligible or have significant impacts on process performance.

Teunter and Flapper (2011) showed that for a remanufacturing system with quality-based core sortation policy, even if expected fractions of each quality levels are known, exact fractions are still uncertain and this uncertainty impacts PrAM decisions. Using a closed-form cost model, uncertainty of quality fractions and associated impacts on sorting decisions are analyzed. It is assumed that a remanufacturer can optimize core acquisition quantity. Hence, a remanufacturer can acquire optimal number of cores without any uncertainty. Optimal product acquisition under uncertain core quality conditions was also investigated by Yang et al. (2014). However, they formulated the core cost assessment problem as a non-linear integer programming model. Remanufacturing lead time is adopted as quality metric such that the higher the remanufacturing lead time, the lower the quality. Since cores were assumed to be collected from brokers on an as-needed basis, uncertainties in quantity and timing are not integrated in the developed linear cost model.

Sutherland and Haapala (2010) studied remanufacturing of diesel engines. By integrating the impacts of product yield, remanufacturing efficiency, transportation cost, and economies of scale, they developed a linear cost model to analyze total annual costs of the facility. Their cost model has several components including a component associated with remanufacturing cost. In their model, it is unknown how core quality condition impacts remanufacturing cost. Also, they used annual output of the facility as a criterion to design facility size. However, uncertainty in return quantity and probability of extreme returns impact facility design decisions was not studied.

This literature review indicates that analyzing the impacts of core quality condition, returns quantity and timing on remanufacturing cost over multiple time periods is still a challenge to existing remanufacturing cost models. This research addresses this challenge by developing a

Stochastic Cost of Remanufacturing Model (SCoRM) where impacts of core quality condition, returns quantity and returns timing on remanufacturing cost are jointly analyzed in normal and extreme scenarios.

## 3. Stochastic Cost of Remanufacturing Model (SCoRM)

In remanufacturing systems core return process is a stochastic process. Generally, EoU cores are returned in random quality condition, random time periods and in batches of random size that can be extreme if cores are acquired passively. This stochastic process is schematically shown in Fig.2.

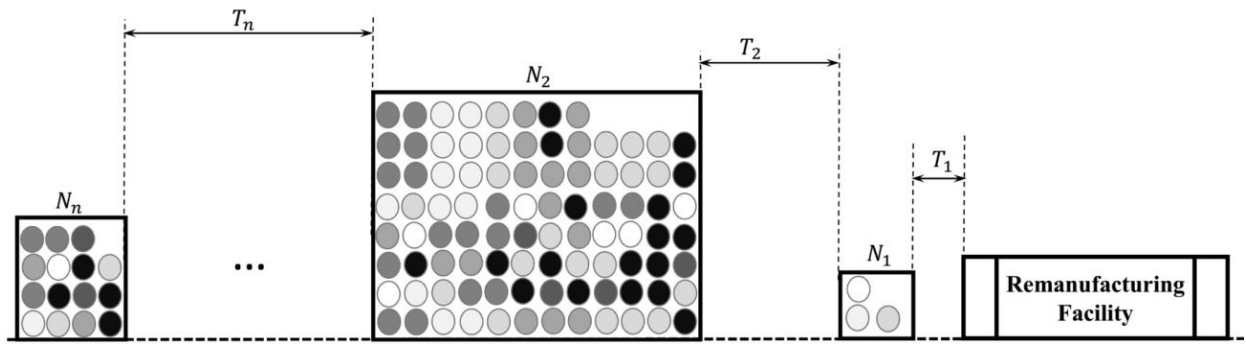

Figure 2. Schematic of core return process

The facility receives cores in batches tagged as $i=1,2,…,n$. Batch size, $N_i$, is a random variable that can be extreme. Inter-arrival time periods are random variables denoted by $T_i$. Core quality is a random variable in [0,1] interval with some distribution function $q(X)$ where $X$ is the random vector denoting quality attributes. SCoRM is developed in three steps: 1) modelling of return process, 2) modelling quality of cores, and 3) joining the models developed in two previous steps by a polynomial cost function. The list of variables and associated descriptions used in development of SCoRM is provided in Table 2.

The following assumptions are made; 1) Batch sizes can be normal or extreme but follow a mixture HPD. 2) Batch return process has Markov property. 3) Following Galbreth and Blackburn (2010), all cores are assumed to be remanufacturable but at different costs. 4) Like Ferguson et al. (2006) remanufacturing cost of a core is assumed to be a polynomial function of core quality condition.

Table 2. Symbols and associated descriptions used in SCoRM

| Symbol | Description |
|---|---|
| $i$ | Batch index |
| $j$ | Core index |
| $T_i$ | Random variable denoting return time period of the $i^{th}$ batch |
| $N_i$ | Batch size that is the number of the cores inside the $i^{th}$ batch |
| $n$ | Total number of returned batches |
| $\mu$ | Mean batch size |
| $\sigma$ | Standard deviation of batch sizes |
| $u$ | Threshold of mixture Hybrid Pareto Distribution (HPD) |
| $\xi$ | Shape factor of mixture HPD |
| $\beta$ | Scale factor of mixture HPD |
| $m$ | Binary random variable denoting batch type that can be normal or extreme |
| $p$ | Probability of extreme return |
| $\mathbf{X}$ | Random vector denoting core quality attributes |
| $q(\mathbf{X}_{ij})$ | Random variable denoting quality of the $j^{th}$ core in the $i^{th}$ batch and $q \in [0,1]$ |
| $c_{ij}$ | Random variable denoting remanufacturing cost of the $j^{th}$ core in the $i^{th}$ batch |

### 3.1. Modeling of core return process

Typically, in remanufacturing applications, core returns have been characterized by Central Tendency (CT) models like Normal distribution and Poisson process where it is assumed that batch sizes, $N_i$, do not hugely deviate from mean batch size, $\mu$. CT models have been extensively used by investigators like Aras et al. (2006), Guide and Van Wassenhove (2006), Galbreth and Blackburn (2006), Teunter and Flapper (2011), and Xiang et al. (2014).

Although normal returns are more frequent, there is solid evidence that cores can be returned in extreme batches. Guide (2006) has described extreme returns as a nuisance to remanufacturers. Though rare, extreme events can be destabilizing and risky (King, and Zeng 2001) and can threaten the sustainment of a remanufacturing system. Despite criticality, extreme returns have been paid less attention in remanufacturing literature. This is probably attributable to rare nature of these events, especially; it is a common practice in CT approaches to exclude the less-likely events from statistical analyses (Montgomery et al., 2009). For triage purposes, Gavidel and Rickli (2015, 2016) modeled extreme returns of industrial valves to a valve shop of a chemical complex. They showed that using Extreme Value (EV) models can complement existing CT models.

In remanufacturing operations, both normal and extreme returns are probable. A Mixture Hybrid Pareto Distribution (HPD) is capable of characterizing both normal and extreme return behavior (Scarrott and Mac Donald, 2012). Mixture HPD is a probabilistic model composed of a truncated

Normal distribution and Generalized Pareto Distribution (GPD) characterizing normal and extreme scenarios, respectively (Carreau et al., 2009). Truncated Normal and GPD are stitched at the threshold of GPD. The threshold value is a criterion to distinguish between normal and extreme events (Coles at al., 2001). Equation 1 presents the Probability Density Function (PDF) of a typical mixture HPD model.

$$HPD(N|\mu,\sigma,u,\xi,\beta) = \begin{cases} \frac{1}{\sigma\sqrt{2\pi}} \exp\left(-\frac{(N-\mu)^2}{2\sigma^2}\right) & ; N < u \\ \begin{cases} \frac{1}{\beta}\left(1+\frac{\xi}{\beta}(N-u)\right)^{-\frac{1}{\xi}-1} & ; \xi \neq 0 \\ \frac{1}{\beta}\exp\left(-\frac{N-u}{\beta}\right) & ; \xi = 0 \end{cases} & ; N \geq u \end{cases} \quad (1)$$

where $N$ is batch size, $\mu$ and $\sigma$ are mean and standard deviation of the truncated Normal distribution and $u$, $\xi$, $\beta$ are threshold, shape, and scale parameters of GPD. Figure 3 presents PDF of a mixture HPD model where the threshold separates normal returns *i.e.*, below-threshold batches from extreme returns *i.e.*, above-threshold batches. In Figure 3, the threshold of the HPD model, depending on the size of batches, classifies the returns into one of the two classes *i.e.*, normal or extreme. Hence, batch type label, *m*, is a random binary variable following Bernoulli distribution described by Eq. 2. Batch type label, *m*, is 0 if return is normal and is 1 if the returned batch is extreme.

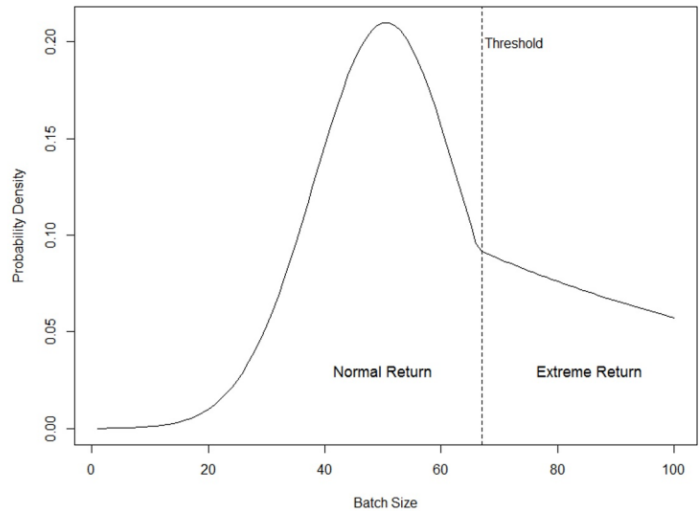

Figure 3. Probability density function of a mixture HPD model

$$B(m,p) = p^m(1-p)^{1-m} \tag{2}$$

where $p$ is the probability of extreme returns. Since returns occur at random time periods (weekly, hourly, and so forth), inter-arrival time periods for extreme returns follow Geometric distribution presented in Eq. 3.

$$G(T_i) = (1-p)^{T_i-1}p, \quad i = 1,2,3,\ldots \tag{3}$$

where $T_i$ is return period for $i^{th}$ batch measured with respect to return period of $(i-1)^{th}$ batch and $G(T_0)=0$. Note that by this formulation, batch return process is a Bernoulli process where batch type, $m$, follows the Bernoulli distribution and batch returns timing follow Geometric distribution. From operational perspective, core return process can also be considered as a Discrete Time Markov Chain (DTMC) with two transition possibilities in the state space *i.e.*, normal or extreme. For more details regarding DTMC process see (Costa et al., 2006).

### 3.2. Assessment of core quality condition

An efficient quality model describing core quality condition, $q$, by using quality attributes vector, $X$, is essential. Quality models can be generated analytically or by data-driven techniques (Van Wassenhove and Zikopoulos, 2009; Gavidel and Rickli, 2017). In this investigation, like Teunter and Flapper (2011), Souza et al. (2002), Ferguson1et al. (2009), and Van Wassenhove and Zikopoulos (2009); it is assumed that core remanufacturing cost and core quality condition are functionally related. Following Ferguson et al. (2009), a polynomial cost function presented in Eq. 4 is adopted.

$$c_{ij} = a_0(1 - q_{ij}^\theta) \tag{4}$$

where $c_{ij}$ is the remanufacturing cost of the $j^{th}$ core in the $i^{th}$ batch, $q_{ij}$ denotes its quality condition, $a_0$ is remanufacturing cost of a core with lowest possible quality, and $\theta$ is the model parameter that can be estimated by process data. For the $i^{th}$ batch with size of $N_i$, batch remanufacturing cost, $C_B$, can be assessed by Eq. 5.

$$C_B = \sum_{j=1}^{N_i} c_j(q(X_j)) \tag{5}$$

Note that $C_B$ is a random variable that depends on the quality of each individual core and size of the batch. For $n$ returned batches, total remanufacturing cost, $C_T$, can be evaluated by Eq. 6.

$$C_T = \sum_{i=1}^{n} \sum_{j=1}^{N_i} c_{ij}\left(q(X_{ij})\right) \tag{6}$$

Remanufacturing costs may vary at normal and extreme conditions. For instance, in extreme scenarios, the facility may hire temporary staff at higher rates to increase remanufacturing capacity to mitigate negative effects of extreme returns. Hence, Eq. 6 can be generally rewritten as:

$$C_T = \sum_{i=1}^{n^0} \sum_{j=1}^{N_i<u} c_{ij}^0\left(q(X_{ij})\right) + \sum_{i=1}^{n^1} \sum_{j=1}^{N_i \geq u} c_{ij}^1\left(q(X_{ij})\right) \tag{7}$$

where $n^0$ and $n^1$ denote number of batches returned in normal and extreme conditions, respectively. Similarly, $c^0$ and $c^1$ denote remanufacturing cost functions for normal and extreme conditions. Finally, applying polynomial cost function presented in Eq. 4, SCoRM can be presented by Eq. 8:

$$C_T = \left(\sum_{i=1}^{n^0} \sum_{j=1}^{N_i<u} a_0^0 \left(1 - q(X_{ij})^{\theta_0}\right)\right) + \left(\sum_{i=1}^{n^1} \sum_{j=1}^{N_i \geq u} a_0^1 \left(1 - q(X_{ij})^{\theta_1}\right)\right) \tag{8}$$

where $a_0^0$ and $\theta_0$, $a_0^1$ and $\theta_1$ are parameters of SCoRM associated with normal and extreme conditions, respectively.

## 4. Case study: remanufacturing of steam traps

In this section, SCoRM is applied to a case study to assess remanufacturing costs of End-of-Use (EoU) steam traps returned to a valve shop in a chemical complex. The underlying core return process is described in section 4.2, and the associated data presented and the SCoRM is generated. SCoRM is used to predict remanufacturing costs and generating the remanufacturing cost paths. Then the prediction accuracy of SCoRM is assessed by using common error measures such as Mean Square Error (MSE) and percent error ($e\%$). Uncertainties present in core quality condition, return quantity and timing propagate and accumulate in the remanufacturing costs. These uncertainties have been analyzed in the case study through bootstrapping. Effects of extreme returns on the remanufacturing cost can be reflected by big jumps on the cost path. The predictive performance of the SCoRM is quantified by using Mean Squared Error (MSE) and percent error ($e\%$). Although MSE and $e\%$ quantify the predictive performance of the SCoRM, it is not known if the MSE and $e\%$ associated with SCoRM are sufficiently low. To address this issue, in the section 4.2, the prediction performance of SCoRM is compared with known predictive algorithms.

### 4.1. Underlying core return process and building SCoRM

In industrial complexes, steam is commonly used as an energy source to heat buildings, process fluids and energize heat transfer equipment. Inherent with the use of steam, are difficulties associated with condensation and accumulation of non-condensable gases in steam networks. Steam traps, type of industrial valves, must be used in steam networks to automatically purge condensate and non-condensable gases. However, a steam trap should never discharge live steam. Such discharges are dangerous and costly (Smith and Mobley, 2003). According to US Department of Energy (DOE), costs of discharging live steam by even smallest steam traps can reach $8,000 per year (US Department of Energy, 2017).Thus, leaking steam traps in a steam network that typically has hundreds to thousands of steam traps will waste millions of dollars per year. To prevent these losses, effective trap maintenance programs are deployed, and accurate cost analysis is critical to these programs.

Steam traps as EoU cores are returned stochastically at random time periods, in batches of random size with random quality conditions. The valve shop must admit all returning cores passively. In this investigation, quality condition of a steam trap is a continuous random variable in [0,1] and is defined in terms of live steam leakage. Such that higher the live steam loss, lower the trap quality (the quality of steam traps with no leakage is 1).

In this complex, a trap maintenance program based on ISO 50001 is deployed and in normal conditions, steam traps return from operational units according to planned Preventive/Predictive Maintenance (PM/PdM) programs. In planned programs, return rate is regulated with processing capacity of the valve shop (35 valves per week) and returns are planned not to be extreme. However, in some rare, unplanned events, like explosions and unexpected overhauls, the quantity of returns can be extreme.

Steam traps return and quality data has been collected from the deployed Computerized Maintenance Management System (CMMS) for *T=81* weeks with totally 1429 steam traps returned in 81 batches. The original steam trap dataset had 29 features but 15 features are relevant to this study. Table 3 presents a list and brief description of these features.

Table 3. Features of the steam trap dataset

| Feature | Description |
| --- | --- |
| Tag Number | An ID to recognize each individual steam trap |
| T | Return period that is also used to identify the returned batches |
| Batch Type | The type of the returning batch that can be normal or extreme |
| Unit | Operational unit of the returned batch there are *12* operational units in this complex |
| Trap Type | Has three levels including Disk, Float, and Thermodynamic |
| Manufacturer | The complex uses products of *7* steam trap manufacturers |
| Application | The functionality of the trap that can be Dripping, Heating, Tracing |
| Con. Size | Is the connection size of the steam trap and has *6* levels |
| Pressure | The operational pressure of the steam trap that can be Low, Medium, and High |
| Wear and Tear | Is *1* if the steam trap has major wears and is *0* otherwise |
| Age | The operation period (*months*) of the steam trap before remanufacturing |
| Cap G. Mat. | Steam trap cap gasket material that has three levels |
| Temp. | Surface temperature of the steam trap in Celsius |
| Leak Rate | Rate of leaking steam measured in kilograms per hour |

An ultrasonic trap monitoring device has been used to measure the leak rate of the steam traps. Figure 4 presents the leak rate of the steam traps and shows that leak rate of steam traps can reach to 35 (kg/hr). Further analysis revealed that, in total, 3051.75 (kg/hr) live steam is wasted in this chemical complex through steam traps leakage.

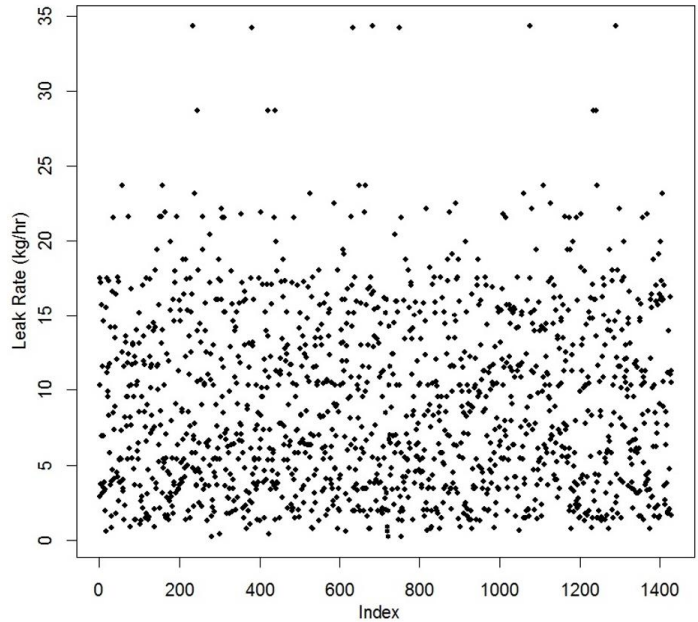

Figure 4. Leak rate of steam traps

Batch size of returns, returns timing, and quality condition (leak rate) are required for SCoRM. The observed remanufacturing cost of steam traps are also required to estimate $\theta_0$ and $\theta_1$ parameters. Figure 5 presents a scatterplot of batch sizes versus return period for *T=81* weekly periods.

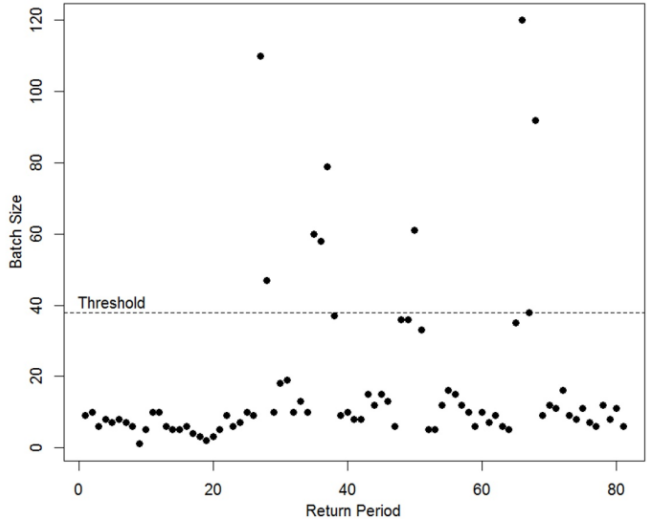

Figure 5. Steam traps returns scatterplot

In Fig. 5, the threshold (*u=38*) is estimated by fitting the mixture Hybrid Pareto Distribution (HPD). The majority of batches are normal and do not exceed this threshold. Parameters of the mixture HPD are estimated by Maximum Likelihood Estimation (MLE) technique and results are as follows. Estimated mean batch size is $\hat{\mu} = 9.82$ where standard deviation of batch size is estimated as $\hat{\sigma} = 22.93$, the threshold of HPD model is $\hat{u} = 38$ where its shape and scale parameters are estimated as $\hat{\xi} = 0.84$ and $\hat{\beta} = 121.75$, respectively. Goodness of Fit (GoF) is tested by conducting Pearson's $\chi^2$ GoF test by following statistical hypotheses:

$$\begin{cases} H_0: Core\ return\ quantity\ (batch\ size)\ follows\ mixture\ HPD \\ H_1: Core\ return\ quantity\ (batch\ size)\ does\ not\ follow\ mixture\ HPD \end{cases}$$

Results of this GoF test are provided in Table 4. It can be statistically inferred that null hypothesis cannot be rejected at common significance levels like 0.01, 0.05, and 0.1.

Table 4. Results of Pearson's $\chi^2$ GoF test

| Test Statistics | $\chi^2$ | Degree of Freedom | p-value |
|---|---|---|---|
| Estimated Value | 2349 | 2320 | 0.3322 |

Figure 6 presents the Probabiltiy Density Function (PDF) of the fitted mixture HPD. The probability parameter of the Bernoulli process is estimated as $\hat{p} = 0.11$. The number of normal and extreme batches are $n^0=72$ and $n^1=9$, respectively.

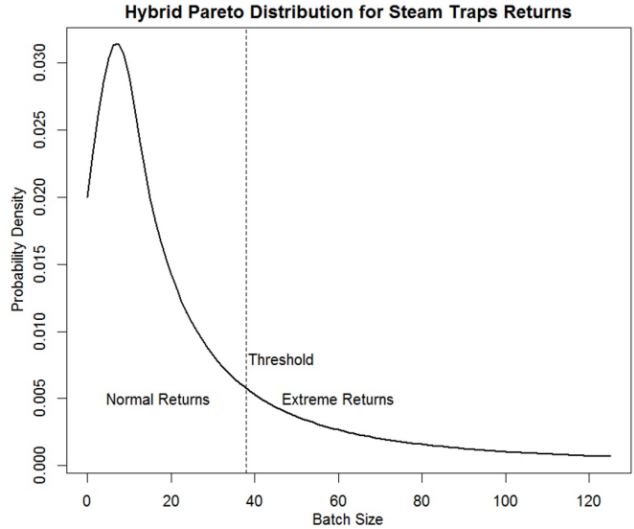

Fidure 6. Mixture HPD model describing return quantity of steam traps

The cost of remanufacturing of a core with minimum quality condition is equal in normal and extreme conditions and $a^0 = a^1 = 500$ (kCurrency). However, power parameter of polynomial cost function, $\theta$, in Eq. 5 is different for normal and extreme scenarios and are estimated as $\hat{\theta}_0=0.64$ and $\hat{\theta}_1=0.76$. To estimate $\hat{\theta}_0$ and $\hat{\theta}_1$, Least Square (LS) method is employed (Christopher, 2016). Figure 7 shows the scatterplot of the observed remanufacturing costs of steam traps versus the corresponding quality condition where the fitted polynomial cost functions are overlaid. Cores returned in normal and extreme return conditions are represented by 0 and 1, respectively.

Figure 7 indicates that for a given core quality condition, remanufacturing costs of cores belonging to extreme batches are generally higher than those belonging to normal returns. In the case of extreme returns, the valve shop uses strategies such as outsourcing remanufacturing tasks and short-time staffing to mitigate/resolve destablizing impacts of extreme returns. These mitigation activities increase the remanufacturing costs compared to normal conditions where the remanufacturing capacity of the valve shop is sufficient to remanufacture all the returning valves and no outsourcing and short-time staffing is required.

From Fig. 7, it can be observed that for a given return type (0/1) and quality condition, remanafacturing costs can be highly inconsistent. These inconsistencies are attributable to imperfections other than leaking. For example, consider two steam traps with the same leak rate and return type, but at different wear conditions. The valve shop, regardless of degree of wear, must treat all the imprefections and restore both of the traps to like-new conditions.

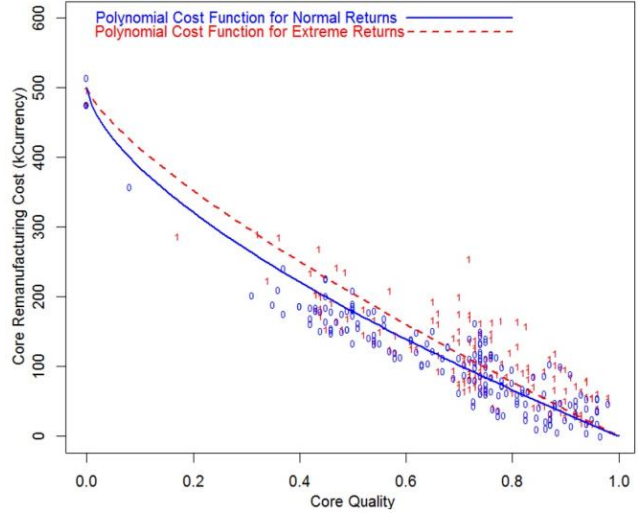

Figure 7. Fitted polynomial cost function for normal and extreme returns

The predicted batch remanufacturing costs, $\hat{C}_{B-SCoRM}$, along with observed costs $C_{B-Obs.}$, harvested from CMMS are reported in Table 5.

In Table 5, $\bar{q}$, is average quality of a batch. Approximately 47% of the total number of returns and 42% of total remanufacturing costs are due to 11% of the batches that have been admitted in the extreme conditions, *i.e., T={ 27, 28, 35, 36, 37, 50, 66, 67, 68}*.

Table 5 also shows that some non-extreme batches like *T={1, 49}*, have higher remanufacturing costs compared to some extreme batches like *T={28, 37, 68}*. So, from this observation, it can be inferred that remanufacturing costs depend on both returns quantity and quality. Further analysis reveals that though *T={1, 49}* are normal but due to the low quality of the cores, the associated remanufacturing costs are higher than remanufacturing costs of batches associated with *T={28, 37, 68}*. This observation re-emphasizes the necessity of joint analysis of these factors in cost evaluations. According to Table 5, the MSE (Christopher, 2016) in predicting batch remanufacturing costs by SCoRM is *MSE=1583*. Note that percent error, *e%*, can be evaluated by Eq. 9.

$$e\% = |\frac{\hat{C}_{T-SCoRM} - C_{T-Obs.}}{C_{T-Obs.}}| \times 100 \qquad (9)$$

where $\hat{C}_{T-SCoRM}$ is the total remanufacturing cost predicted by SCoRM and $C_{T-Obs.}$ is the total observed remanufacturing cost. Considering the total remanufacturing cost estimated by SCoRM and the observed total remanufacturing cost (see Table 5), using Eq. 9, the percent error is estimated as *e%= 1.55%*. Note that SCoRM underestimates the total remanufacturing cost.

Table 5. Results of applying SCoRM to steam trap dataset

| T | N | m | $\bar{q}$ | $C_{B-Obs.}$ | $\hat{C}_{B-SCoRM}$ | T | N | m | $\bar{q}$ | $C_{B-Obs.}$ | $\hat{C}_{B-SCoRM}$ |
|---|---|---|---|---|---|---|---|---|---|---|---|
| 1 | 9 | 0 | 0.09 | 2140 | 2177 | 42 | 8 | 0 | 0.65 | 395 | 375 |
| 2 | 10 | 0 | 0.67 | 1176 | 1225 | 43 | 15 | 0 | 0.75 | 736 | 725 |
| 3 | 6 | 0 | 0.47 | 292 | 357 | 44 | 12 | 0 | 0.78 | 628 | 563 |
| 4 | 8 | 0 | 0.93 | 65 | 99 | 45 | 15 | 0 | 0.79 | 751 | 711 |
| 5 | 7 | 0 | 0.91 | 56 | 47 | 46 | 13 | 0 | 0.83 | 312 | 308 |
| 6 | 8 | 0 | 0.38 | 644 | 641 | 47 | 6 | 0 | 0.5 | 269 | 243 |
| 7 | 7 | 0 | 0.91 | 92 | 83 | 48 | 36 | 0 | 0.47 | 1355 | 1346 |
| 8 | 6 | 0 | 0.69 | 93 | 83 | 49 | 36 | 0 | 0.33 | 2551 | 2531 |
| 9 | 1 | 0 | 0.64 | 153 | 133 | 50 | 61 | 1 | 0.48 | 5165 | 5156 |
| 10 | 5 | 0 | 0.85 | 106 | 105 | 51 | 33 | 0 | 0.88 | 664 | 620 |
| 11 | 10 | 0 | 0.83 | 184 | 164 | 52 | 5 | 0 | 0.9 | 80 | 96 |
| 12 | 10 | 0 | 0.82 | 179 | 156 | 53 | 5 | 0 | 0.8 | 106 | 136 |
| 13 | 6 | 0 | 0.86 | 125 | 120 | 54 | 12 | 0 | 0.72 | 631 | 558 |
| 14 | 5 | 0 | 0.34 | 438 | 339 | 55 | 16 | 0 | 0.55 | 701 | 721 |
| 15 | 5 | 0 | 0.44 | 321 | 337 | 56 | 15 | 0 | 0.73 | 490 | 496 |
| 16 | 6 | 0 | 0.36 | 421 | 385 | 57 | 12 | 0 | 0.94 | 218 | 209 |
| 17 | 4 | 0 | 0.68 | 158 | 163 | 58 | 10 | 0 | 0.97 | 158 | 151 |
| 18 | 3 | 0 | 0.64 | 112 | 107 | 59 | 6 | 0 | 0.75 | 186 | 239 |
| 19 | 2 | 0 | 0.95 | 35 | 20 | 60 | 10 | 0 | 0.46 | 590 | 641 |
| 20 | 3 | 0 | 0.66 | 156 | 144 | 61 | 7 | 0 | 0.75 | 286 | 232 |
| 21 | 5 | 0 | 0.98 | 121 | 101 | 62 | 9 | 0 | 0.94 | 71 | 94 |
| 22 | 9 | 0 | 0.78 | 156 | 179 | 63 | 6 | 0 | 0.99 | 52 | 35 |
| 23 | 6 | 0 | 0.33 | 335 | 342 | 64 | 5 | 0 | 0.89 | 93 | 83 |
| 24 | 7 | 0 | 0.72 | 830 | 830 | 65 | 35 | 0 | 0.79 | 1116 | 1188 |
| 25 | 10 | 0 | 0.86 | 171 | 193 | 66 | 120 | 1 | 0.78 | 3670 | 3636 |
| 26 | 9 | 0 | 0.96 | 56 | 38 | 67 | 38 | 1 | 0.53 | 2607 | 2602 |
| 27 | 110 | 1 | 0.74 | 3123 | 3101 | 68 | 92 | 1 | 0.92 | 1200 | 1258 |
| 28 | 47 | 1 | 0.78 | 1105 | 1008 | 69 | 9 | 0 | 0.49 | 445 | 477 |
| 29 | 10 | 0 | 0.45 | 653 | 629 | 70 | 12 | 0 | 0.89 | 120 | 173 |
| 30 | 18 | 0 | 0.82 | 169 | 111 | 71 | 11 | 0 | 0.94 | 163 | 128 |
| 31 | 19 | 0 | 0.42 | 1193 | 1163 | 72 | 16 | 0 | 0.84 | 462 | 385 |
| 32 | 10 | 0 | 0.79 | 335 | 315 | 73 | 9 | 0 | 0.9 | 78 | 83 |
| 33 | 13 | 0 | 0.89 | 182 | 171 | 74 | 8 | 0 | 0.96 | 127 | 80 |
| 34 | 10 | 0 | 0.64 | 410 | 395 | 75 | 11 | 0 | 0.78 | 243 | 262 |
| 35 | 60 | 1 | 0.75 | 1788 | 1713 | 76 | 7 | 0 | 0.81 | 121 | 174 |
| 36 | 58 | 1 | 0.38 | 3169 | 3162 | 77 | 6 | 0 | 0.81 | 401 | 304 |
| 37 | 79 | 1 | 0.97 | 1032 | 1015 | 78 | 12 | 0 | 0.95 | 48 | 104 |
| 38 | 37 | 0 | 0.5 | 1975 | 1882 | 79 | 8 | 0 | 0.57 | 370 | 334 |
| 39 | 9 | 0 | 0.52 | 879 | 824 | 80 | 11 | 0 | 0.97 | 267 | 232 |
| 40 | 10 | 0 | 0.48 | 852 | 800 | 81 | 6 | 0 | 0.47 | 241 | 270 |
| 41 | 8 | 0 | 0.5 | 706 | 693 | | | | | | |

Uncertainties in core quality condition, return quantity and timing propagate and accumulate in remanufacturing costs. The accumulated uncertainties and associated variations have been assessed by using bootstrapping technique according to Friedman, et al. (2001). Figure 8 shows the results of this analysis where the original predicted cost path (blue) generated by SCoRM and *3000* bootstrapped cost paths (gray) are superimposed. Expected remanufacturing cost path (dashed) is also included in Fig. 8 for comparison purposes.

Figure 8 shows that, uncertainties originating from core quality condition, return quantity and timing can make the cost predictions highly uncertain. Note that these uncertainties are pronounced in the presence of extreme returns that are reflected as big jumps on the cost path. Bootstrapping analysis reveals that at the best scenario, total remanufacturing cost can be as low as 27,702 and in the worst case it can be as high as 101,360 where expected total cost is 54,270. Knowing the extent of variations of predicted remanufacturing costs assists decision makers to make more informed decisions. For example, considering the range of variations in process cost, a manager can reserve sufficient financial resources to cover remanufacturing costs in the case of extreme returns such that operational destabilizations and can be mitigated/resolved.

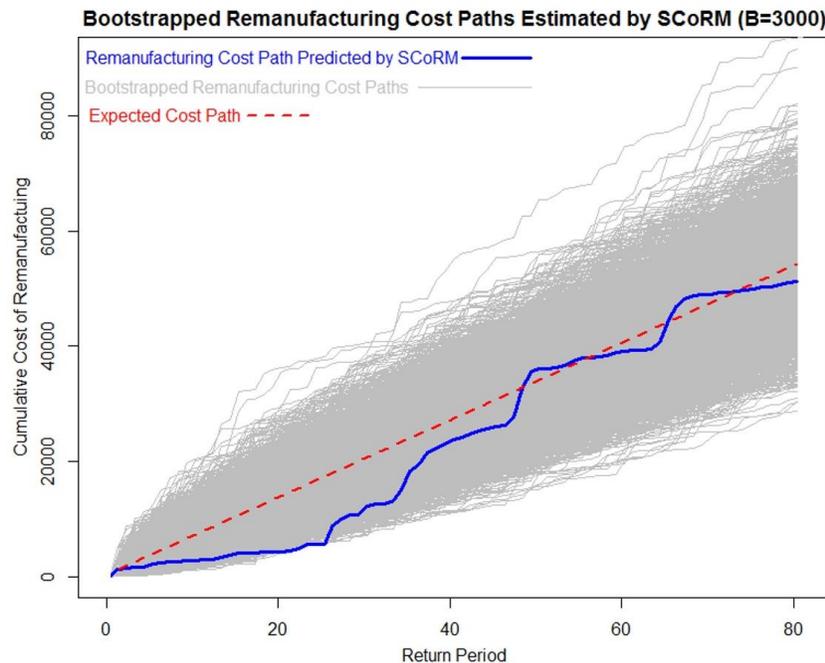

Figure 8. Predicted (blue) and bootstrapped (gray) cost paths generated by SCoRM

*4.2. Prediction performance of SCoRM*

Data-driven predictive models have been applied to range of re/manufacturing applications like process optimization (Sadati et al., 2018), reliability and sustainability assessments (Fard et al., 2015; Roostaei and Zhang, 2016; Moradi and Huang, 2016). In this section, data-driven predictive algorithms have been used to conduct a comparative analysis. For this reason, predictive performance of SCoRM is compared with Generalized Additive Model (GAM), Support Vector Machine-Sequential Minimal Optimization (SVM-SMO), Artificial Neural Network- Multilayer Perceptron (ANN-MLP), and Random Forest (RF) algorithms. For details regarding these algorithms see Friedman, et al. (2001), Smola and Schölkopf (2004), Christopher (2016), and Ho (1998), respectively. The ZeroR algorithm is used as a baseline for comparing the prediction approaches (Katz, 2000). This comparison is conducted to provide deeper insights regarding predictive performance of SCoRM.

In section 4.1, the predictive performance of SCoRM is evaluated and the mean square and percent errors are estimated as *MSE=1583* and *e%=1.55%*. Although predictive performance of SCoRM is quantified in this way, deeper insights into its predictive performance are provided by comparing to aforementioned predictive algorithms. To generate the data-driven predictive models, data harvested from the CMMS along with a ten-fold cross-validation scheme according to Friedman, et al. (2001) is utilized. The input factors including T, Unit, Trap Type, Manufacturer, Application, Con. Size, Pressure, Wear and Tear, Age, Cap G. Mat., Temp., and Leak Rate are used to predict the remanufacturing cost of the steam trap as the output factor. Then, the predicted remanufacturing cost of the steam traps were used to calculate the remanufacturing cost of the associated batches. Finally, remanufacturing cost of the batches were summed cumulatively to generate the cost paths. The predicted cost paths are mapped versus return periods in Fig 9. For comparison purposes, the observed cost path and the cost path predicted by SCoRM (blue) are superimposed on Fig. 9.

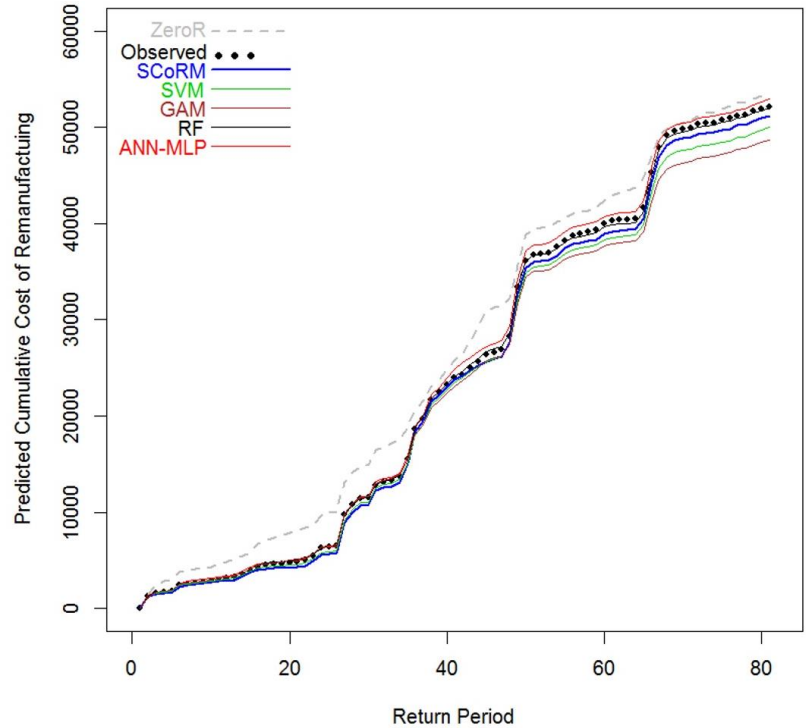

Figure 9. Observed and predicted cost paths

The ZeroR algorithm sets an upper bound for the error value of predictive models such that no predictive model should have error greater than ZeroR (Katz, 2000). Table 6 presents the MSE and $e\%$ of the predictive models in predicting batch remanufacturing costs of 81 returned batches. SCoRM, like other sophisticated predictive algorithms, outperforms ZeroR algorithm in terms of MSE, thus possesses prediction performance.

Table 6. Predictive performance validation of SCoRM

|  | ZeroR | GAM | SVM-SMO | SCoRM | ANN-MLP | RF |
|---|---|---|---|---|---|---|
| *MSE* | 13422 | 1788 | 1730 | 1583 | 1212 | 586 |
| *e%* | 3.15 | 10.43 | 4.06 | 1.55 | 1.44 | 0.87 |

Note that RF, SCoRM, SVM, and GAM are underestimating the total remanufacturing cost (see Fig.9) while ZeroR and ANN-MLP are overestimating the total remanufacturing cost. Table 6 indicates that RF with the minimum MSE, has the best prediction performance. SCoRM outperforms ZeroR, GAM and SVM-SMO but is outperformed by RF and ANN-MLP. Even though ANN-MLP and RF outperform SCoRM in remanufacturing cost prediction, SCoRM has advantages in remanufacturing applications. SCoRM is specifically developed by considering the requirements and models existing in the literature of remanufacturing. Second, SCoRM is an analytical model that can be used for simulation purposes while data-driven models need data to be generated. From technical perspective ANN-MLP and RF have major limitations. According to Christopher (2016) prediction process and relation between variables in ANN-MLP is like black-box and cannot be explained easily, while SCoRM illustrates the relationship between key remanufacturing parameters.

## 5. Conclusion and extensions

Uncertainties in core quality condition, return quantity and timing jointly propagate and accumulate in process cost and complicate cost assessments. In this paper, the joint effects of core quality condition, return quantity, and timing on remanufacturing cost under normal and extreme return conditions is studied. A multivariate model called the Stochastic Cost of Remanufacturing Model (SCoRM) was developed in order to incorporate joint effects of returns quality, quality, and timing condition into remanufacturing cost assessments. Within SCoRM, a mixture Hybrid Pareto Distribution (HPD) is used to characterize normal and extreme core acquisition scenarios. By using the threshold of HPD model, it has been shown that for remanufacturing systems with passive core returns, the return process can be modeled as a Bernoulli process. SCoRM is used to assess remanufacturing costs of steam traps returning to a valve shop of a chemical complex. Remanufacturing costs are assessed by cost paths that illustrate remanufacturing costs as a function of batch arrivals. It is shown that extreme returns can be reflected as big jumps on the cost path. Also, variations of SCoRM in predictive tasks is assessed by bootstrapping technique. Results of this variation analysis shows that in the presence of extreme returns, remanufacturing costs can be highly uncertain.

Through a comparative analysis, the predictive performance of SCoRM is compared with the Generalized Additive Model (GAM), Support Vector Machine-Sequential Minimal Optimization (SVM-SMO), Artificial Neural Network- Multilayer Perceptron (ANN-MLP), and Random Forest (RF). The comparative analysis conducted by using two performance assessment measures, Mean Square Error (MSE) and percent error (*e%*). Results of this analysis indicates that SCoRM's performance is comparable. SCoRM outperforms GAM and SVM-SMO algorithms but is outperformed by ANN-MLP and RF. In this paper, it is assumed that core return timing is a random discrete variable following a Geometric distribution, however, future work will model core return timing as a continuous distribution.

# References

Aras, N., Verter, V. and Boyaci, T., 2006. Coordination and priority decisions in hybrid manufacturing/remanufacturing systems. *Production and Operations Management*, *15*(4), pp.528-543.

Bhamra, T. and Hon, B., 2004. *Design and Manufacture for Sustainable Development 2004*. John Wiley & Sons.

Cai, X., Lai, M., Li, X., Li, Y. and Wu, X., 2014. Optimal acquisition and production policy in a hybrid manufacturing/remanufacturing system with core acquisition at different quality levels. *European Journal of Operational Research*, *233*(2), pp.374-382.

Carreau, J., Naveau, P. and Sauquet, E., 2009. A statistical rainfall-runoff mixture model with heavy-tailed components. *Water resources research*, *45*(10).

Christopher, M.B., 2016. *PATTERN RECOGNITION AND MACHINE LEARNING*. Springer-Verlag New York.

Choi, T.M. and Cheng, T.E. eds., 2011. *Supply chain coordination under uncertainty*. Springer Science & Business Media.

Coles, S., Bawa, J., Trenner, L. and Dorazio, P., 2001. *An introduction to statistical modeling of extreme values* (Vol. 208). London: Springer.

Costa, O.L.V., Fragoso, M.D. and Marques, R.P., 2006. *Discrete-time Markov jump linear systems*. Springer Science & Business Media.

D'Adamo, I. and Rosa, P., 2016. Remanufacturing in industry: advices from the field. *The International Journal of Advanced Manufacturing Technology*, *86*(9-12), pp.2575-2584.

Energy Efficiency and Renewable Energy, US Department of Energy 2017, accessed 1 December 2017, https://energy.gov/sites/prod/files/2013/11/f4/webinar_steamtrap_2010_0605_0.pdf.

Fadeyi, J.A., Monplaisir, L. and Aguwa, C., 2017. The integration of core cleaning and product serviceability into product modularization for the creation of an improved remanufacturing-product service system. *Journal of Cleaner Production*, *159*, pp.446-455.

Fard, M.J., Ameri, S. and Hamadani, A.Z., 2015. Bayesian approach for early stage reliability prediction of evolutionary products. In *Proceedings of the International Conference on Operations Excellence and Service Engineering* (pp. 361-371).

Ferguson, M., Guide, V.D., Koca, E. and Souza, G.C., 2009. The value of quality grading in remanufacturing. *Production and Operations Management*, *18*(3), pp.300-314.

Friedman, J., Hastie, T. and Tibshirani, R., 2001. *The elements of statistical learning* (Vol. 1, pp. 337-387). New York: Springer series in statistics.

Galbreth, M.R. and Blackburn, J.D., 2006. Optimal acquisition and sorting policies for remanufacturing. *Production and Operations Management*, *15*(3), pp.384-392.

Gavidel, S.Z. and Rickli, J.L., 2018. End-of-Use Core Triage in Extreme Scenarios Based on a Threshold Approach. *arXiv preprint arXiv:1801.03093*.

Gavidel, S.Z. and Rickli, J.L., 2017. Quality assessment of used-products under uncertain age and usage conditions. *International Journal of Production Research*, *55*(23), pp.7153-7167.

Gavidel, S.Z. and Rickli, J.L., 2015. Triage as a core sorting strategy in extreme core arrival scenarios. *Journal of Remanufacturing*, *5*(1), p.9.

Guide Jr, V.D.R., 2000. Production planning and control for remanufacturing: industry practice and research needs. *Journal of operations Management*, *18*(4), pp.467-483.

Guide, V.D.R. and Wassenhove, L.N., 2001. Managing product returns for remanufacturing. *Production and operations management*, *10*(2), pp.142-155.

Guide, V.D.R. and Wassenhove, L.N., 2006. Closed-loop supply chains: an introduction to the feature issue (part 1). *Production and Operations Management*, *15*(3), pp.345-350.

Hauser, W.M. and Lund, R.T., 2008. *Remanufacturing: operating practices and strategies: perspectives on the management of remanufacturing businesses in the United States*. Department of Manufacturing Engineering, Boston University.

Ho, T.K., 1998. The random subspace method for constructing decision forests. *IEEE transactions on pattern analysis and machine intelligence*, *20*(8), pp.832-844.

Ilgin, M.A. and Gupta, S.M., 2012. *Remanufacturing modeling and analysis*. CRC Press.

Jiang, Z., Zhou, T., Zhang, H., Wang, Y., Cao, H. and Tian, G., 2016. Reliability and cost optimization for remanufacturing process planning. *Journal of cleaner production*, *135*, pp.1602-1610.

Katz, A.R., 2000. Image Analysis and Supervised learning in the automated Differentiation of White Blood cells from Microscopic Images.

King, G. and Zeng, L., 2001. Logistic regression in rare events data. *Political analysis*, *9*(2), pp.137-163.

Moradi-Aliabadi, M. and Huang, Y., 2016. Multistage optimization for chemical process sustainability enhancement under uncertainty. *ACS Sustainable Chemistry & Engineering*, *4*(11), pp.6133-6143.

Montgomery, D.C., Runger, G.C. and Hubele, N.F., 2009. *Engineering statistics*. John Wiley & Sons.

Organisation for Economic Co-operation and Development, 2001. *Extended Producer Responsibility: A Guidance Manual for Governments*. OECD Publishing.

Robotis, A., Boyaci, T. and Verter, V., 2012. Investing in reusability of products of uncertain remanufacturing cost: the role of inspection capabilities. *International Journal of Production Economics*, *140*(1), pp.385-395.

Roostaei, J. and Zhang, Y., 2017. Spatially explicit life cycle assessment: opportunities and challenges of wastewater-based algal biofuels in the United States. *Algal Research*, *24*, pp.395-402.

Sadati, N., Chinnam, R.B. and Nezhad, M.Z., 2018. Observational data-driven modeling and optimization of manufacturing processes. *Expert Systems with Applications*, *93*, pp.456-464.

Scarrott, C. and MacDonald, A., 2012. A review of extreme value threshold es-timation and uncertainty quantification. *REVSTAT–Statistical Journal*, *10*(1), pp.33-60.

Smith, R. and Mobley, R.K., 2003. *Industrial machinery repair: best maintenance practices pocket guide*. Butterworth-Heinemann.

Smola, A.J. and Schölkopf, B., 2004. A tutorial on support vector regression. *Statistics and computing*, *14*(3), pp.199-222.

Steinhilper, R., 1998. *Remanufacturing: the ultimate form of recycling* (Vol. 44). Stuttgart: Fraunhofer-IRB-Verlag.

Sutherland, J.W., Jenkins, T.L. and Haapala, K.R., 2010. Development of a cost model and its application in determining optimal size of a diesel engine remanufacturing facility. *CIRP annals*, *59*(1), pp.49-52.

Teunter, R.H. and Flapper, S.D.P., 2011. Optimal core acquisition and remanufacturing policies under uncertain core quality fractions. *European Journal of Operational Research*, *210*(2), pp.241-248.

Yang, C.H., Wang, J. and Ji, P., 2015. Optimal acquisition policy in remanufacturing under general core quality distributions. *International Journal of Production Research*, *53*(5), pp.1425-1438.